\documentclass[twocolumn,prl,amsmath,amssymb,pre]{revtex4}
\usepackage{bm}
\usepackage{xcolor}
\usepackage{graphicx}

\newcommand{\w}{\omega}

\newcommand{\B}{\mbox{\tiny B}}

\newcommand{\Q}{\mbox{\tiny Q}}

\renewcommand{\d}{{\rm d}}

\newcommand{\ti}{\tilde}

\newcommand{\la}{\langle}
\newcommand{\ra}{\rangle}

\newcommand{\be}{\begin{equation}}
\newcommand{\ee}{\end{equation}}
\newcommand{\bea}{\begin{eqnarray}}
\newcommand{\eea}{\end{eqnarray}}
\newcommand{\bsube}{\begin{subequations}}
\newcommand{\esube}{\end{subequations}}
\newcommand{\Eq}[1]{Eq.\,(\ref{#1})}
\newcommand{\Eqs}[1]{Eqs.\,(\ref{#1})}
\newcommand{\Fig}[1]{Fig.\,\ref{#1}}

\newcommand{\comments}[1]{}

\begin{document}
\title{On the lower bound of the Heisenberg uncertainty product in the Boltzmann states}

\author{Yao Wang} %
\affiliation{
Hefei National Research Center for Physical Sciences at the Microscale,
	University of Science and Technology of China, Hefei, Anhui 230026, China}
	
\date{\today}

\begin{abstract}
The uncertainty principle lies at the heart of quantum mechanics, as it describes the fundamental trade-off between the precision of position and momentum measurements. In this work, we study the quantum particle in the Boltzmann states and derive a refined lower bound on the product of 
$\Delta x$ and 
$\Delta p$. Our new bound is expressed in terms of the ratio between 
$\Delta x$ and the thermal de Broglie wavelength, and provides a valuable tool for characterizing thermodynamic precision. We apply our results to the Brownian oscillator system, where we compare our new bound with the well-known Heisenberg uncertainty principle. Our analysis shows that our new bound offers a more precise measure of the thermodynamic limits of precision.
\end{abstract}
\maketitle

\paragraph{Introduction.}
The uncertainty principle is a cornerstone of quantum mechanics, expressing the fundamental limitation on our ability to simultaneously measure certain pairs of physical observables \cite{Hei49}. The most well--known version of the uncertainty principle is the Heisenberg uncertainty relation, which places a lower bound on the product of the standard deviations of position and momentum for a quantum particle \cite{Hei27172}.
It reads
\be \label{HUR}
\Delta x \Delta p\geq \frac{\hbar}{2},
\ee
where $\Delta x \equiv \sqrt{\la \delta \hat x\delta \hat x\ra} $, $\Delta p \equiv \sqrt{\la \delta \hat p\delta \hat p\ra} $ and $\hbar$ is the reduced Planck constant.
Equation (\ref{HUR}) holds for any quantum pure or mixed state, giving a universal lower bound of the Heisenberg uncertainty product $\Delta x \Delta p$.
It is also usually interpreted as some upper limit to the precision of our measurements in microscopic world and inspires  a lot of studies and discussions on this thread; see e.g. Ref\, \cite{Bus07155,Rud1238003,Roz12100404,Bao205658,Has232828}.

In recent years, there has been growing interest in understanding the 
precision of observables 
in quantum  thermodynamic regimes \cite{Lee18032119,Bra18090601,Has19110602,Tim19090604,Sei19176,Has19062126,Gua19033021,Hor2015,Van22140602,Fu22024128,Kam23L052101}.
Especially, when in the equilibrium Boltzmann state, the system density operator is then proportional to $\exp(-\beta \hat H)$ with $\beta\equiv 1/(k_B T)$ being the inverse temperature and $\hat H$ the Hamiltonian.
 One natural question arises:  Can the lower bound of the Heisenberg uncertainty product $\Delta x \Delta p$ be refined when the system is constrained in the Boltzmann equilibrium state? 

On the other hand, when the thermodynamic effects are involved at finite temperatures,
the thermal de Broglie wavelength,
\be 
\lambda_{\rm th}\equiv\sqrt{\frac{2\pi\hbar^2}{mk_{B}T}},
\ee
can be roughly viewed the average de Broglie wavelength of particles, where $m$ is the mass of the particle \cite{Hua87,Cha87}.
It describes the wave--like nature of particles at certain temperatures. 
At higher temperatures, particles have greater kinetic energy, which leads to shorter thermal de Broglie wavelengths.
Therefore, the thermal de Broglie wavelength is crucial for predicting and analyzing the behavior of particles under different temperatures. For example, when $\lambda_{\rm th}\sim n^{-1/3}$ with $n$ the particle number density, the quantum effect is anticipated to be prominent \cite{Hua87,Cha87}.
So, another  question comes: How shall we connect $\lambda_{\rm th}$ to the thermodynamic uncertainty of precision?

\begin{figure}[h]
\includegraphics[width=0.85\columnwidth]{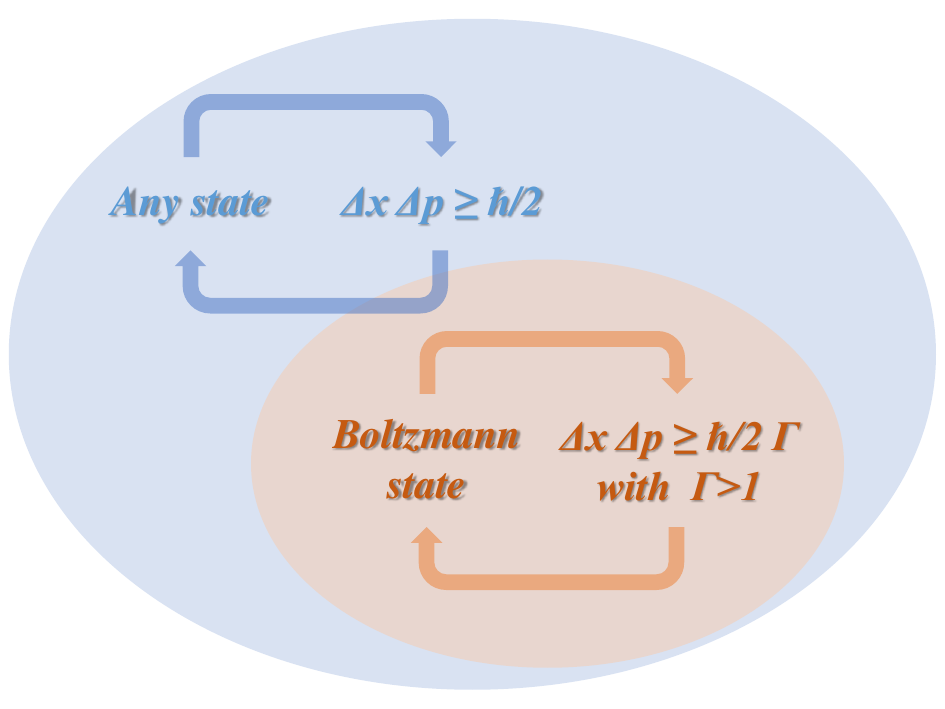}
\caption{An illustrative figure of Heisenberg lower bound (in light blue) [cf.\,\Eq{HUR}] versus Boltzmann lower bound (in orange) [cf.\,\Eq{BUR}].}\label{fig1}
\end{figure}

In this work, we report on the discovery of a refined lower bound, named as the Boltzmann lower bound, of the Heisenberg uncertainty product in the Boltzmann states.
The uncertainty relation is then expressed  as
\be \label{BUR}
\Delta x\Delta p\geq \frac{\hbar}{2}\times\Gamma \Big(\frac{1}{4\pi r^2}\Big)\ \ \text{with}\ \  r\equiv \Delta x/\lambda_{\rm th}.
\ee 
Here, $\Gamma(x)> 1$ is a constructed function reading
\be \label{gammax}
\Gamma(x)=\frac{g^{-1}(x)}{x}, 
\ee 
where $g^{-1}$ is the inverse function of $g(x)=x\tanh(x/2)$.
In \Eq{BUR}, $r$ is defined as
 the ratio of $\Delta x$ to the thermal de Broglie wavelength.
Generally, \Eq{BUR} supplies us with a lower bound that is encoded with both the quantum ($\hbar$) and thermodynamic ($\lambda_{\rm th}$) features; see \Fig{fig1}.

The remainder of this paper is organized as follows.
Next, we present the step--by--step derivation of the Boltzmann lower bound in \Eq{BUR}.
The demonstration and discussion are then carried out on the Brownian oscillator system, including the comparison with the celebrated lower bound of Heisenberg. 
 Finally, we summarize the paper.

\paragraph{Boltzmann lower bound.}

In this section, we will present the  step--by--step derivation of \Eq{BUR}.
The methodology here is
closely related to that exploited in deriving thermodynamic uncertainty relations \cite{Str22}.

\emph{Step 0: Notations and preliminary knowledge.}
For later use, we denote 
\be \label{ct}
c(t)\equiv \la \delta \hat x(t)\delta \hat x(0)\ra,
\ee
where the average $\la(\,\cdot \,)\ra$ runs over the Boltzmann state $\propto\exp(-\beta \hat H)$. 
Evidently, 
\be\label{c0} 
c(t=0)=(\Delta x)^2.
\ee

 In the non-relativistic scenario where $\hat p$ is only involved in the kinetic energy term as $\hat p^2/(2m)$, we further have \cite{Yan05187}
\be  \label{dc0}
 \dot c(0)=\frac{1}{m}\la \delta \hat p\delta\hat  x\ra=-\frac{1}{m}\la \delta \hat x\delta \hat p\ra=-\frac{i\hbar} {2m},
\ee 
and 
  \be \label{ddc0}
  \ddot c(0)=-\bigg(\frac{\Delta p}{m}\bigg)^2.
  \ee
In deriving \Eqs{dc0} and (\ref{ddc0}),  we have used the differential time--reversal relation, which reads \cite{Yan05187}
\be \label{tdiff}
\la \dot {\hat A}(t)\hat B(0)\ra= -\la \hat A(t)\dot {\hat B}(0)\ra
\ee
for any operator $\hat A$ and $\hat B$ in the equilibrium states.
Set $\hat A=\hat B=\delta\hat x$ and $t=0$ in \Eq{tdiff}, by further noting $\delta\dot{\hat x}=i[\hat H,\delta\hat x]=\hat p/m$ and $\la \hat p\ra=0$, and we have
\be
\la \delta\hat p\delta\hat x\ra=-\la \delta\hat x\delta\hat p\ra.
\ee
This, together with the relation$\la[\delta\hat x,\delta\hat p]\ra\equiv \la\delta\hat x\delta\hat p\ra-\la\delta\hat p\delta\hat x\ra =i\hbar$, gives rise to \Eq{dc0}.
Similarly, we can easily obtain \Eq{ddc0} by setting $\hat A=\delta\hat p$ and $\hat B=\delta\hat x$.

\emph{Step 1: Detailed balance and positivity.}
 Let us start the derivation of \Eq{BUR} with the detailed balance relation reading \cite{Cal5134, Yan05187}
\be \label{db}
  \frac{C(-\w)}{ C(\w)}=e^{-\beta \hbar \w},
\ee
where [cf.\,\Eq{ct}]
\be \label{ft} 
C(\w)\equiv\frac{1}{2}\int_{-\infty}^{\infty}\!{\rm d}t\,e^{i\w t} c (t)
\ee
is  real and positive definite, i.e.
\be 
\forall \w,\ \  [C(\w)]^{\ast}=C(\w) \ \ \text{and}\ \  C(\w)\geq 0.
\ee

Therefore, one may introduce a probability distribution $P(\w)$ over the frequency domain as
\be \label{P}
P(\w)\equiv \frac{C(\w)}{\pi (\Delta x)^2},\ \ \ \ \w\in(-\infty,\infty).
\ee
The normalization factor $\pi (\Delta x)^2$ can be inferred from
the inverse transform of \Eq{ft}, 
\be \label{ct}
c(t)=\frac{1}{\pi}\int_{-\infty}^{\infty}\!{\rm d}\w\,e^{-i\w t}C(\w),
\ee
by setting $t=0$.
Equation (\ref{db}) then directly leads to 
\be \label{db4}
P(-\w)=e^{-\beta \hbar \w}P(\w).
\ee
This is the equation to be used in the next step.

\emph{Step 2: The distribution $Q(\w)$.}
To proceed, we further introduce a  distribution  defined only on the interval $\w\in[0,\infty)$ as \cite{Str22}
\be 
Q(\w)=(1+e^{-\beta \hbar \w})P(\w).
\ee
By using \Eq{db4}, it is easy to verify $Q(\w)\geq 0$ and 
\[
\int_{0}^{\infty}\!{\rm d}\w\,Q(\w)=1.
\]
This means $Q(\w)$ is also a normalized distribution over the definition domain.
Hereafter, we  denote expectation values of any
function $f(\w\geq 0)$ with respect to $P(\w)$ by $\la f(\w) \ra$, while  denote that to $Q(\w)$ by $\la f(\w) \ra_{\Q}$. 

The first and second moments of $\w$ with respect to $P(\w)$ and $Q(\w)$ are then connected as
\be \label{1stm}
\beta \hbar \la \w\ra=\la g(\beta \hbar \w)\ra_{\Q}
\ee
and
\be \label{2ndm}
 \la \w^2 \ra=\la \w^2 \ra_{\Q}.
\ee
In \Eq{1stm},
\be \label{g}
g(x)=x\tanh(x/2).
\ee

\emph{Step 3: Convexity and inequality.}
It is easy to verify that $g(\sqrt{x})$ is monotonically increasing and concave for $x\geq 0$ \cite{Str22}. Therefore, its inverse function $k^2$, with $k=g^{-1}$, must be convex .

Then according to Jensen's inequality \cite{Jen06175}, we have
\be\label{key}
\begin{split}
\la \w^2 \ra&\overset{(\ref{2ndm})}{=\joinrel=}\la \w^2 \ra_{\Q}=\frac{\la k^2[g(\beta \hbar \w)] \ra_{\Q}}{(\beta \hbar)^{2}}
\\ &
\geq \frac{k^2[\la g(\beta \hbar \w) \ra_{\Q}]}{(\beta \hbar)^{2}}
\overset{(\ref{1stm})}{=\joinrel=}\frac{k^2(\beta  \hbar  \la \w\ra)}{(\beta \hbar)^2}.
\end{split}
\ee
The second equality in the first line is deduced from $[k g(\sqrt{x})]^2=x$, by setting $x=(\beta \hbar \w)^2$.
The ``$\geq$'' in the second line is the Jensen inequality for the convex function $k^2$.
Finally, we obtain
\be \label{eq21}
\la \w^2 \ra\geq \frac{k^2(\beta  \hbar  \la \w\ra)}{(\beta \hbar)^2}
\ee
in this step.

\emph{Step 4: Relation to $\Delta x$ and $\Delta p$.}
As inferred from \Eqs{P} and (\ref{ct}), together with \Eqs{c0}--(\ref{ddc0}), we know that 
\be 
\la  \w\ra=\frac{\hbar}{2m(\Delta x)^2}
\ee
and
\be 
\la  \w^2\ra=\Big(\frac{\Delta p}{m\Delta x}\Big)^2.
\ee
Then after some simple algebram \Eq{eq21} gives rise to \Eq{BUR}, with 
\be \label{gammax}
\Gamma(x)=\frac{k(x)}{x}=\frac{g^{-1}(x)}{x}.
\ee
Here, $g(x)$ is as defined in \Eq{g}.

\begin{figure}[h]
\includegraphics[width=1\columnwidth]{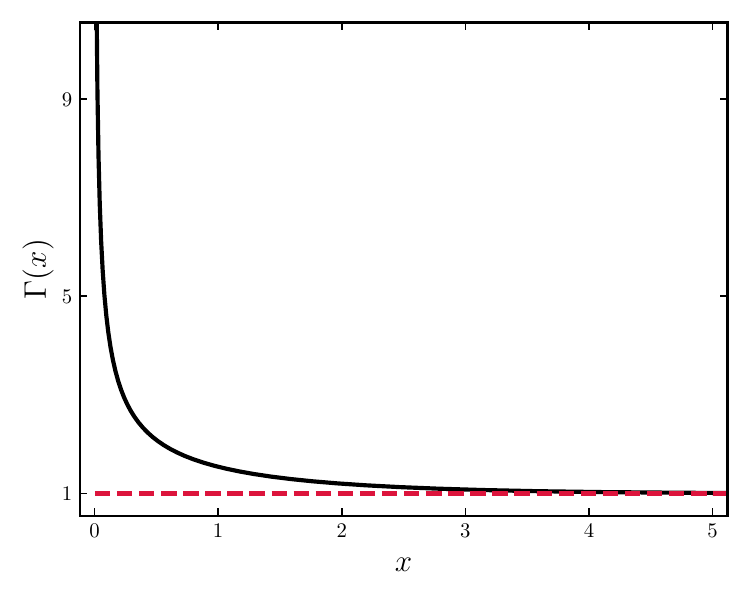}
\caption{The image of function $\Gamma(x)$; see \Eq{gammax} with \Eq{g}.}\label{fig2}
\end{figure}

It is easy to see $\Gamma(x)> 1$ for all $x>0$.
To prove this, one may just verify $g(x) < x$. This is obvious since $\tanh(x/2)< 1$ for all $x>0$.
Here we plot the image of the function $\Gamma(x)$ in \Fig{fig2} for readers' reference.
As shown in the \Fig{fig2}, when $x\rightarrow \infty$, $\Gamma(x)\rightarrow 1$.

\paragraph{Numerical demonstrations.}
As a illustration, we consider  a Brownian oscillator system with the Hamiltonian in the Calderia--Leggett form \cite{Cal83587,Cal83374,Wei12,Yan05187},
\be \label{HM_DBO}
\begin{split}
 \hat H&= \left(\frac{\hat p^2}{2m} + \frac{1}{2}m\Omega^2 \hat x^2\right)
\\ &\quad
    + \sum_j \left[\frac{p_j^2}{2m_j}
         +\frac{1}{2} m_j\omega_j^2
     \Big(x_j-\frac{c_j}{m_j\omega_j^2}\hat x\Big)^2\right] .
\end{split}
\ee
It involves a vibrational mode $\hat x$ and $\hat p$ and its coupling to
a solvent bath,
\be 
h_{\B}=\sum_j \Big(\frac{p_j^2}{2m_j} +\frac{1}{2} m_j\omega_j^2x_j^2\Big).
\ee

The  response  of the vibrational mode is defined as 
\be\label{chiqqt}
   \chi(t) \equiv \frac{i}{\hbar}\la[\hat x(t), \hat x(0)]\ra,
\ee
where we have denoted
\be \label{ope_env}
 \hat x(t) \equiv e^{iH t/\hbar}\hat x (0)e^{-iH t/\hbar}.
\ee
In the frequency domain, its reads \cite{Yan05187}
\be \label{chiwhar}
\begin{split}
 \ti\chi(\omega)\equiv \int^{\infty}_{0}\!\!\d t\, e^{i\w t} \chi(t)= \frac{1/m}{\Omega^2 - \omega^2
     - i\omega\ti\gamma(\omega)}.
\end{split}
\ee
Here, the bath--induced friction function is
\be 
\ti \gamma(\w)\equiv \int_{0}^{\infty}\!\!{\rm d}t\,e^{i\w t} \gamma(t)
\ee
 with 
\be
     \gamma(t) =
  (1/m)\sum_j c_j^2/(m_j\omega_j^2)\cos(\omega_jt)
\ee
being the classical frictional function. 
In simulation, we adopt the Drude model for the bath, as
\be 
\ti \gamma(\w)=\frac{i\eta \Omega}{\w+i\zeta}
\ee
where $\eta$ and $\zeta$ are two parameters characterizing the coupling strength and damping rate, respectively.

\begin{figure}[h]
\includegraphics[width=1\columnwidth]{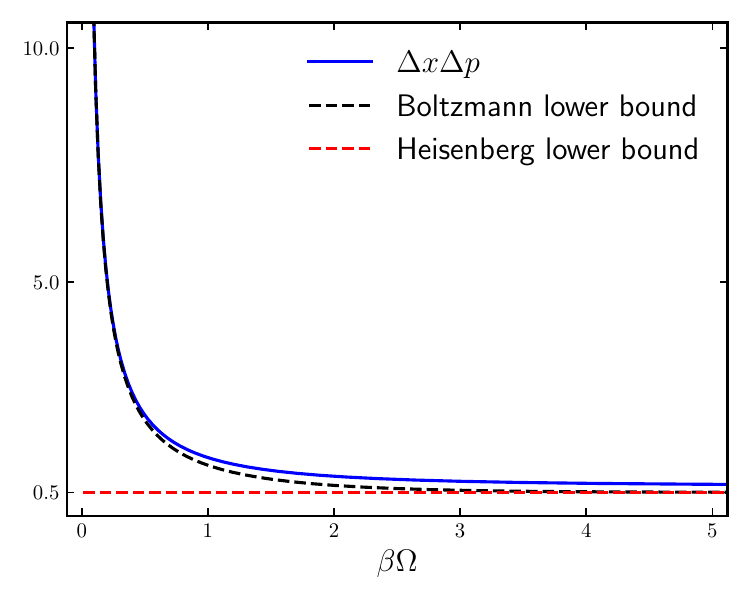}
\caption{The uncertainty $\Delta x\Delta p$ of the Brownian oscillator system in the Boltzmann states at different temperatures. We set $\hbar=k_{B}=1$, $m=\Omega^{-1}$, and $\ti \gamma(\w)=i\eta \Omega/(\w+i\zeta)$ with $\eta=\zeta=10\Omega$.}\label{fig3}
\end{figure}

The $C(\w)$ in \Eq{ft} is related to $\ti \chi(\w)$ via the fluctuation--dissipation theorem reading \cite{Cal5134, Yan05187}
\be 
C(\w)=\frac{\hbar\,{\rm Im}\ti \chi(\w)}{1-e^{-\beta\w}},
\ee
or equivalently [cf.\,\Eq{ct}],
\be \label{eq34}
c(t)=\frac{\hbar}{\pi}\int_{-\infty}^{\infty}\!\!{\rm d}\w\,e^{-i\w t}\frac{{\rm Im}\ti \chi(\w)}{1-e^{-\beta\w}}.
\ee
From \Eq{eq34}, we can compute the $\Delta x$ and $\Delta p$, and also generator the Boltzmann lower bound.
We obtain them numerically at different temperatures and plot the data in \Fig{fig3}.

As shown in \Fig{fig3}, our numerical results support our findings. Besides, we can observe that the Boltzmann lower  bound coincides with that of Heisenberg, $\hbar/2$,  at extremely low temperatures ($\beta\Omega \gg 1$); see the black versus red dash lines in this regime. 
This is in line with our physical perception: the thermodynamic effect tapers off when the temperature tends to zero.

Howerver, it is evident that \Eq{BUR} gives rise to a much more refined lower bound at relatively high temperatures ($\beta\Omega\sim 1$); see the blue solid versus black dash lines.
Here, it is the constraint of the thermodynamic canonical distribution that further reduces the precision expressed via the  the Heisenberg uncertainty product $\Delta x \Delta p$.

\paragraph{Summary.}
To summarize, in this work we investigate the Heisenberg uncertainty product in the context of quantum particles in Boltzmann states. 
A refined lower bound for this product is derived, which is expressed as a function of the ratio between the standard deviation of position  and the thermal de Broglie wavelength.
We demonstrate the Boltzmann lower bound on the Brownian oscillator system and compare the results with the traditional Heisenberg uncertainty principle. 

Overall, the Boltzmann lower bound represents a new tool for characterizing the thermodynamic precision, and opens up new avenues for future researches in quantum thermodynamics and related fields. 
It may inspire studies exploring the uncertainty relations in other contexts.
For example, exploring the impact of thermal effects on the Heisenberg uncertainty principle may shed light on the complex interactions between quantum mechanics and thermodynamics \cite{Lan54643,Elc57161,Koc81380,Mig202471,Koy22014104}.
From a practical perspective, the refined lower bound developed in this work could be useful in fields such as nanotechnology, where precise measurement of position and momentum is crucial \cite{Gar04,Cle101155,Soa14825}.
It could also provide insights into the behavior of complex systems such as biological molecules, where thermal effects are often important \cite{Lee071462,Che09241,Lam1310}.
It is anticipated the finding in this work would become an important ingredient
in a wide range of fields.

Besides, in a recent analysis based on the thermodynamic considerations, it has been shown that the Heisenberg uncertainty relation is deformed in the quantum gravity regime \cite{Buo22136818}. 
Particularly, the lower bound of the Heisenberg uncertainty product gets increased for positive values of the deformation parameter \cite{Buo22136818}. 
It is attractive to establish relationship between the lower bound deformation with the quantum gravity  and that with the thermodynamic considerations presented in this work.
This may help understand the gravity--thermodynamic conjecture, originally formulated for black holes in \cite{Jac951260}, which establishes a profound connection between gravity and thermodynamic effects.

\begin{acknowledgements}
Support from the National Natural Science Foundation of China (No.\ 22103073) and the  USTC New Liberal Arts Fund (No.\  FSSF-A-230110) is gratefully acknowledged.
\end{acknowledgements}
%


\end{document}